# Noise Suppression for CRP Gathers Based on Self2Self with Dropout

Fei Li, Zhenbin Xia, Dawei Liu, Xiaokai Wang, *Member, IEEE,* Wenchao Chen, Juan Chen, and Leiming Xu

*Abstract*—Noise suppression in seismic data processing is a crucial research focus for enhancing subsequent imaging and reservoir prediction. Deep learning has shown promise in computer vision and holds significant potential for seismic data processing. However, supervised learning, which relies on clean labels to train network prediction models, faces challenges due to the unavailability of clean labels for seismic exploration data. In contrast, self-supervised learning substitutes traditional supervised learning with surrogate tasks by different auxiliary means, exploiting internal input data information. Inspired by *Self2Self with Dropout*, this paper presents a self-supervised learning-based noise suppression method called Self-Supervised Deep Convolutional Networks (SSDCN), specifically designed for Common Reflection Point (CRP) gathers. We utilize pairs of Bernoulli-sampled instances of the input noisy image as surrogate tasks to leverage its inherent structure. Furthermore, SSDCN incorporates geological knowledge through the normal moveout correction technique, which capitalizes on the approximately horizontal behavior and strong self-similarity observed in useful signal events within CRP gathers. By exploiting the discrepancy in self-similarity between the useful signals and noise in CRP gathers, SSDCN effectively extracts self-similarity features during training iterations, prioritizing the extraction of useful signals to achieve noise suppression. Experimental results on synthetic and actual CRP gathers demonstrate that SSDCN achieves high-fidelity noise suppression.

## I. INTRODUCTION

FIELD seismic data are susceptible to noise caused by external environmental factors and exploration equipment, leading to a detrimental impact on the quality of seismic data imaging and reservoir prediction accuracy. Consequently, noise suppression has emerged as a critical research area in seismic data processing.

Researchers worldwide have dedicated efforts to exploring and proposing various effective methods for suppressing noise in seismic data. These methods can be broadly categorized into traditional and deep learning-based techniques, based on the research methodologies employed.

Traditional noise suppression methods for seismic data involve constructing mathematical models that rely on explicit knowledge of the physical mechanisms underlying the data. Mathematical optimization techniques are then employed to solve these models. These traditional methods can be broadly categorized into three groups based on their implementation approaches: filtering [1]–[3], low-rank matrix decomposition [4], [5], and noise suppression methods utilizing sparse representation [6], [7]. Although traditional noise suppression methods have shown promising results in practical applications, they still face two significant challenges. Firstly, many of these methods rely on artificial assumptions and explicit prior information about the physical background, which inherently limits their ability to capture the high complexity and nonlinear relationships within data. Secondly, traditional methods often require manual parameter tuning during the model-solving process, necessitating constant adjustment of empirical values to account for data variations.

In recent years, Convolutional Neural Networks (CNNs) have gained significant attention and success in computer vision. Consequently, researchers have explored applying deep learning techniques in seismic data processing, yielding promising results. Among them, DnCNN [8] has attracted considerable interest and has successfully attenuated various types of noise, such as random noise [9], surface waves [10], desert noise [11], multiple waves [12], and scattering noise [13]. The aforementioned methods highlight the effectiveness of supervised deep learning in noise suppression for seismic data. Deep neural networks possess strong adaptability and nonlinear feature extraction capabilities, making them well-suited for this denoising task. Another key advantage of deep learning lies in its flexibility, as it does not require extensive domain knowledge or specialized expertise. It is worth noting that the efficacy of supervised learning-based noise suppression methods heavily relies on the quality of seismic data labels, as a high-quality dataset can significantly enhance network denoising performance.

In contrast to natural image denoising, a major challenge for seismic exploration is the inability to obtain clean labels for building training sample pairs, preventing further improvements in seismic data denoising. As a result, unsupervised learning-based denoising methods have garnered significant attention since they do not rely on clean labels for network training. Liu et al. [14] introduced GCN, an unsupervised deep neural network that leverages the inherent self-similarity of seismic data. GCN employs a decoder

Manuscript received July *, 2023; revised July, *, 2023 and July *, 2023; accepted **** **, ****. This work was supported by the National Natural Science Foundation of China under Grant 41974131. (Corresponding author: Dawei Liu.)

Fei Li, Juan Chen, and Leiming Xu is with Exploration and Development Research Institute, Petrochina Changqing Oilfield Co., Xi'an, 710018, China (e-mail: lifei@petrochina.com.cn; chenj1_cq@petrochina.com.cn; xulm1_cq@petrochina.com.cn).

Zhenbin Xia, Dawei Liu, Xiaokai Wang, and Wenchao Chen are with the School of Information and Communications Engineering, Xi'an Jiaotong University, Xi'an 710049, China (e-mail: 1040412138@qq.com; liudawei2015@stu.xjtu.edu.cn; xkwang@xjtu.edu.cn; wencchen@xjtu.edu.cn).

structure that can extract multi-scale self-similarity features, prioritizing the reconstruction of useful signals during the iterative process. Quan et al. [15] proposed *Self2Self with Dropout*, a denoising network for a single natural image, which effectively suppresses random noise by exploiting the information within the input image from pairs of Bernoulli-sampled instances. Building upon this, Wang et al. [16] applied *Self2Self with Dropout* to seismic data denoising and successfully suppressed random noise in post-stack seismic data. Sun et al. [17] presented N2STL, an unsupervised learning paradigm for suppressing random noise in 2D seismic data. N2STL utilizes the constructed J-invariant function to learn seismic data information and introduces transfer learning to reduce training time. Consequently, it effectively suppresses random noise in prestack and post-stack seismic data.

This paper presents a Self-Supervised Deep Convolutional Network (SSDCN) method specifically designed for noise suppression in Common Reflection Point (CRP) gathers. Inspired by the *Self2Self with Dropout* approach, SSDCN utilizes multiple training samples derived from Bernoulli-sampled instances of single noisy seismic images to establish optimization objectives. Notably, CRP gathers, after undergoing normal moveout (NMO) correction, exhibit a distinct difference in self-similarity between the useful signals and noise. The useful signal events are characterized by approximate horizontal behavior and pronounced self-similarity. Leveraging its inherent capability to extract self-similarity features, SSDCN prioritizes the extraction of useful signals during the training iterations, resulting in effective noise suppression.

## II. METHOD

### A. Self-Supervised Denoising Principle

Seismic data can be represented as a linear combination of useful signals and noise as follows:

$$y = x + n, \quad (1)$$

where $y$ denotes noisy seismic data, $x$ stands for useful signals, and $n$ signifies noise.

Deep neural networks based on supervised learning usually use Mean Squared Error (MSE) loss to optimize the network model. This optimization ensures that the network output value approximates the desired label value as closely as possible, thereby enhancing the network accuracy in estimating useful signals:

$$\min_{\theta} \frac{1}{N} \sum_{i=1}^{N} \|F_\theta(y_i) - x_i\|_2^2, \quad (2)$$

where $F_\theta(\cdot)$ symbolizes the network, $\theta$ represents the network parameters, $y_i$ and $x_i$ denote the $i$ th input sample and its corresponding label, respectively, and $N$ is the total number of samples in the training dataset.

Seismic exploration encounters a practical issue in acquiring completely clean seismic data labels due to the unpredictability of underground structures. Consequently, directly using (2) as the objective function for network optimization becomes problematic. Inspired by *Self2Self with Dropout*, we turn to Bernoulli sampling. This technique allows us to generate surrogate supervised training pairs from a single CRP gather. Specifically, we use a mask consisting solely of zeros and ones to construct these Bernoulli sampling data pairs, as shown in the following mathematical expressions:

$$\hat{y}_i \coloneqq m_i \odot y, \quad (3)$$
$$\bar{y}_i \coloneqq (1 - m_i) \odot y, \quad (4)$$

where $m_i$ signifies the mask used for the $i$ th Bernoulli sampling. The probability of preserving keeping each pixel point is $\rho \epsilon (0,1)$, and $\odot$ symbolizes the Hadamard product.

The sampled $\hat{y}_i$ differs from the original seismic data $y$, but it still retains most of the valuable information of $y$. More importantly, the sampled signals $\hat{y}_i$ still exhibit self-similarity. After repeating Bernoulli sampling $N$ times and constructing the Bernoulli-sampled dataset $\{(\hat{y}_i, \bar{y}_i)\}_{i=1}^{N}$, we can transform the learning objective of the entire network as follows:

$$\min_{\theta} \sum_{i=1}^{N} \|(1 - m_i) \odot (F_\theta(\hat{y}_i) - \bar{y}_i)\|_2^2. \quad (5)$$

The objective function reaches its minimum when $F_\theta(\hat{y}_i) \to y$. By designing the network with a multi-scale structure, it can extract multi-scale self-similarity features. When the network reconstructs the seismic data, it prioritizes the extraction of useful signals with high self-similarity. Let $\|\cdot\|_{m_i}^2 = \|(1 - m_i) \odot \cdot\|_2^2$. Then, (5) can be simplified as:

$$\min_{\theta} \sum_{i=1}^{N} \|F_\theta(\hat{y}_i) - \bar{y}_i\|_{m_i}^2. \quad (6)$$

As demonstrated in the original *Self2Self* paper [17], if the noise meets the criteria of being independent and having a zero mean, we can perform several Bernoulli sampling experiments, and the optimization of (6) is equivalent to

$$\min_{\theta} \left( \sum_{i=1}^{N} \|F_\theta(\hat{y}_i) - x\|_{m_i}^2 + \sum_{i=1}^{N} \|\sigma\|_{m_i}^2 \right), \quad (7)$$

where $\sigma$ is the standard deviation of the noise $n$. Equation (7) indicates that identical results can be achieved whether we use Bernoulli sampling to construct training samples from the CRP gather itself, then train the network using Bernoulli sampling data $(\hat{y}_i, \bar{y}_i)$, or directly train the network using $(\hat{y}_i, x)$ containing true labels of actual seismic data. Consequently, we opt for (6) as the optimization objective for our proposed self-supervised network. After the training process, we determine the final denoising result by averaging the outcomes of multiple experiments

$$x' = \frac{1}{N} \sum_{i=1}^{N} F_{\theta_i}(\hat{y}_i), \quad (8)$$

where $x'$ denotes the estimated clean seismic data. By using the average value as the final result, we can enhance the network accuracy in reconstructing useful signals.

### B. Network Architecture

Fig. 1 illustrates the SSDCN architecture proposed in this study. Its overall network architecture resembles UNet [18] and features a multi-scale structure. With its multi-scale structure, SSDCN prioritizes the extraction of useful signals exhibiting self-similarity during iterations. Consequently, an appropriate number of iterations can be chosen to enable the network to extract useful signals, thereby achieving noise suppression.

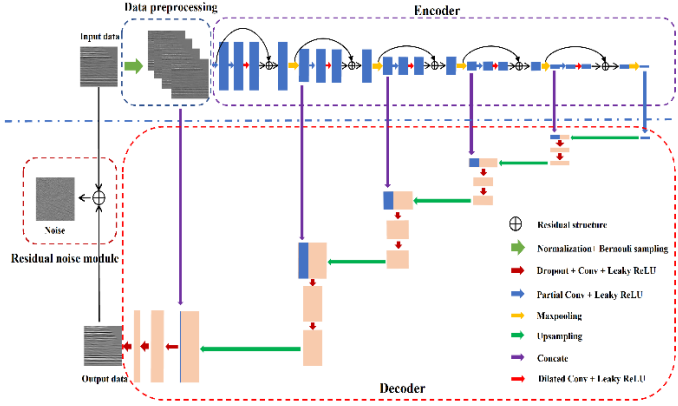

Fig. 1. Network architecture of our SSDCN.

SSDCN comprises four primary components: the data processing module, encoder, decoder, and noise separation module. The data processing module primarily involves data normalization and Bernoulli sampling. The Bernoulli sampling positions are randomized with retention probability $\rho \epsilon (0,1)$, which can be set to 0.7 in the experiment of the next section.

The second component, the encoder module, condenses and extracts features from the preprocessed seismic data through five encoding stages. To enhance feature extraction, we employ dilated convolution to broaden the receptive field of SSDCN, setting a dilation rate of three. Additionally, we integrate a residual learning block to avert network degradation and boost performance. Each encoding stage concludes with a down-sampling process using a maximum pooling layer of size 2×2 and a stride of two, which halves the size of the output feature map relative to the input. After passing through five encoding stages, the feature maps are compressed to 1/32 of the original seismic data size.

The third component, the decoder, incrementally maps the high-level semantic features extracted by the encoder back into clean seismic data. Symmetrical to the encoder, it comprises five decoding stages. Each decoding stage gradually restores the size and resolution of the feature map through upsampling and merges the multi-scale features extracted by the encoder and decoder using skip connections, finally matching the size of the original seismic data. This step-by-step process recovers the target details and improves the reconstruction of the original seismic data. After five rounds of upsampling, the decoder ultimately reconstructs the clean seismic data.

The last component is the residual noise separation module. This module calculates the residual between the original seismic data and the decoder output to identify the removed noise. It implicitly uses prior information with a zero mean as a regularization term to constrain the separated noise (see (7)). This approach prevents damage to useful signals and enhances their fidelity.

## III. DATA EXAMPLES

### A. Synthetic Data Example

To validate the effectiveness of our method, we first conduct experiments on synthetic seismic data examples. As depicted in Fig. 2(a), the clean seismic data consists of 9 gathers with 40 traces, and its useful signals are represented as hyperbolic reflection events. Each single gather has 500 sampling time points, with a sampling time interval of 4 ms. We added random noise to the clean seismic data to synthesize the noisy seismic data shown in Fig. 2(b), resulting in a signal-to-noise ratio (SNR) of –3.77 dB. After NMO correction, the useful signals, which are approximately horizontal events, exhibit good self-similarity, as shown in Fig. 2(c).

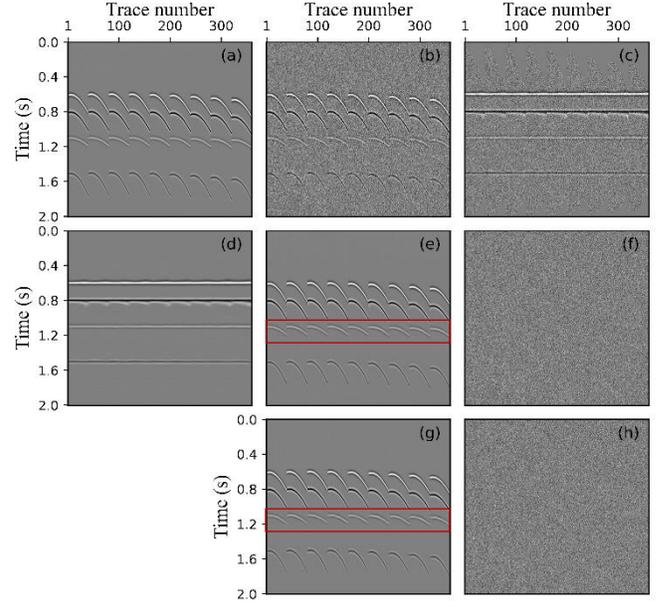

Fig. 2. Denoising results of synthetic seismic data using SSDCN. (a) Original clean data. (b) Noisy data (-3.77 dB). (c) Noisy data after NMO. (d) Denoised by SSDCN. (e) Denoised results after reversing NMO (17.36 dB). (f) Noise removed by SSDCN. (g) Denoised results (14.25 dB) by SSDCN applied to (b). (h) Noise removed from (b).

For comparison, we apply our SSDCN method to suppress noise on the seismic data depicted in Fig. 2(b) and Fig. 2(c). Fig. 2(e) and Fig. 2(f) respectively display the denoising results (with an SNR of 17.36 dB) and the corresponding noise removed by our method on the synthetic data example after NMO. A significant improvement in Fig. 2(e) is evident without compromising useful information. In a similar vein, Fig. 2(g) and Fig. 2(h) respectively present the denoising results (with an SNR of 14.25 dB) and the corresponding noise removed by our method on the noisy data without NMO. While these results are also promising, the SNR value is lower than the previous one. Significantly, flattened events display a high degree of self-similarity. This characteristic aligns with our network's ability to extract multi-scale self-similarity, thereby further boosting the denoising performance of our method. This conclusion is further substantiated when comparing the areas within the red boxes in Fig. 2(e) and Fig. 2(g). After NMO, the structure of useful signals simplifies, facilitating learning by the SSDCN. The final reconstructed useful signals are more consistent in Fig. 2(e), demonstrating that SSDCN is highly effective at recovering useful signals that exhibit self-similarity.

*B. Field CRP Gather Example*

To further validate the effectiveness of our method, we conduct experiments on field seismic data. We use CRP gathers obtained from Eastern China as an example to demonstrate the ability of our encoder-decoder structured network to extract useful signals with self-similarity features. Each CRP gather contains 200 traces with 3001 sampling time points at a 2 ms sampling interval. Fig. 3 displays the one series of denoising results by SSDCN over different iterations. When the number of iterations is small, the network first extracts the overall structure of useful signals. As the number of iterations increases, more details are reconstructed. However, with further increases in the number of iterations, noise will eventually be restored as well.

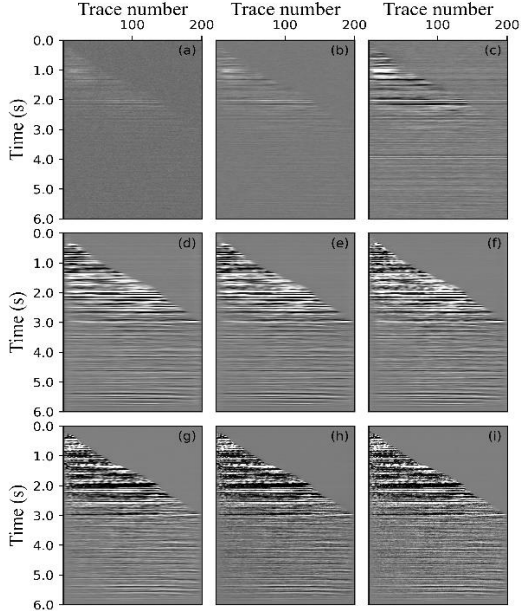

Fig. 3. The denoising results by SSDCN with different iterations. (a)-(i) are results of 100, 500, 1000, 5000, 7000, 10000, 15000, 20000, 30000 iterations, respectively.

The events of the useful signals in the CRP gathers are approximately horizontal, resulting in a high degree of self-similarity. Fig. 4 presents the F-K spectrum corresponding to Fig. 3(a)–(i). The energy of these useful signals is primarily concentrated in the low-frequency and low-wavenumber region of the F-K domain, typically below 75 Hz. Conversely, noise exhibits broadband characteristics, not only overlapping the useful signal region but also containing high-frequency energy. Notably, the proposed network preferentially learns low-frequency features. As the number of iterations increases, it progressively reconstructs more high-frequency components. After approximately 7000 iterations, SSDCN can reconstruct the desired clean seismic data to achieve successful noise suppression. However, if the number of iterations is further increased, noise begins to reappear in the denoising results, as shown in Fig. 4. Therefore, setting an optimal number of iterations is crucial for SSDCN to effectively map noisy field seismic data to clean seismic data.

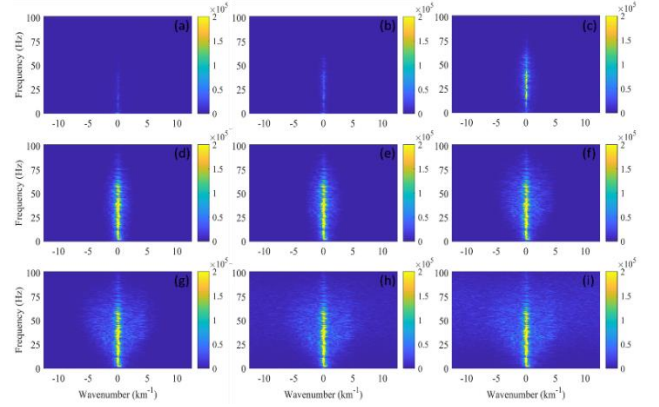

Fig. 4. F-K spectra corresponding to denoising results by SSDCN across different iterations. (a)-(i) are results of 100, 500, 1000, 5000, 7000, 10000, 15000, 20000, 30000 iterations, respectively.

Furthermore, we employ an additional CRP gather from the same dataset to verify the applicability of our method on a larger scale. This gather shares the same size and sampling rate as the previous one. For comparison, we employ F-X filtering and GCN methods, with the denoising results of all three methods presented in Fig. 5. The red amplified area in Fig. 5(a) clearly shows that the events of useful signals are significantly disrupted by background noise, which includes a substantial amount of random noise and a small quantity of coherent noise. F-X filtering effectively eliminates noise from the noisy seismic data, but a minor amount of noise remains, as illustrated in the red amplified region in Fig. 5(b). Conversely, both the GCN and SSDCN enhance the continuity and smoothness of the reconstructed useful signals, demonstrating superior noise suppression effectiveness. Notably, our SSDCN method not only excels in suppressing the noise but also effectively preserves the structure of the useful signals in the shallow part, indicating high fidelity.

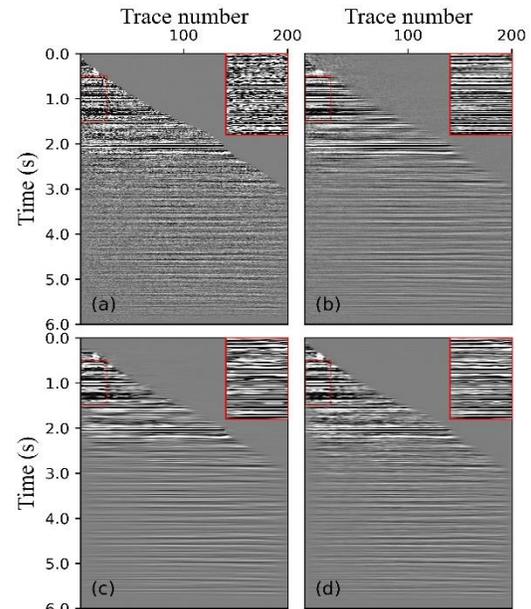

Fig. 5. Comparisons of denoising results with different methods for field CRP gather. (a) Noisy data. (b) Denoised by F-X filtering. (c) Denoised by GCN. (d) Denoised by SSDCN.

Additionally, we calculate the corresponding noise difference for these three methods, as depicted in Fig.-6. It is evident that F-X filtering, GCN, and SSDCN all effectively eliminate the background noise from the original gather. However, F-X filtering distorts the boundaries of the useful signals, causing some damage, as indicated by the red arrow in Fig. 6(a). In contrast, the noise gather generated by our SSDCN method can hardly visualize any residual useful signals, demonstrating high fidelity. Upon examining the red enlarged areas in Fig. 6(a)–(c), it is clear that, besides random noise, these three denoising methods also possess some suppression capability for coherent noise with linear structures.

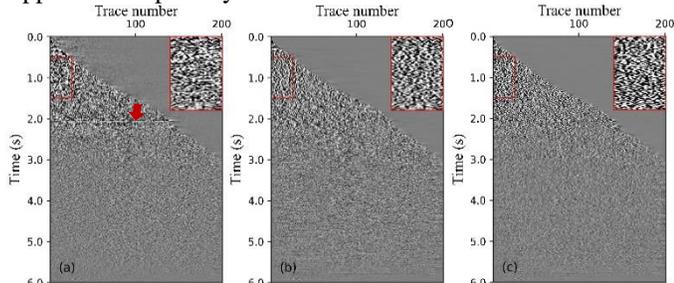

Fig. 6. Difference between Fig 5(a) and Fig 5(b-d). (a) Removed noise by F-X filtering. (b) Removed noise by GCN. (c) Removed noise by SSDCN.

*C. Post-stack CRP Gather Example*

We conduct a stacking process on 250 CRP gathers and subsequently apply SSDCN to denoise the post-stack seismic data, as depicted in Fig. 7(a). Despite stacking being a powerful denoising technique, the stacked section remains significantly disrupted by noise. Fig. 7(b) and Fig. 7(c) display the denoised results of SSDCN and the separated noise, respectively. Our method evidently eliminates a considerable amount of random noise and some coherent noise in the tilt direction, enhancing the continuity of the useful signals. In Fig. 7(c), the absence of any apparent stratigraphic structure in the noise difference section suggests that our method inflicts minimal damage on the useful signals, thus maintaining high fidelity.

V. CONCLUSION

We introduce a self-supervised learning-based noise suppression method for CRP gathers without needing clean labels. We construct surrogate tasks to extract supervised information by conducting multiple Bernoulli sampling experiments on the input CRP gather. By incorporating NMO correction, which introduces geological velocity information, we further enhance the denoising capability of our method. The effectiveness of the SSDCN method is demonstrated through synthetic data and both prestack and poststack CRP gathers, showing its superiority over other denoising methods such as F-X filtering and GCN. Future research could focus on optimizing the number of iterations and enhancing the denoising ability for coherent noise.

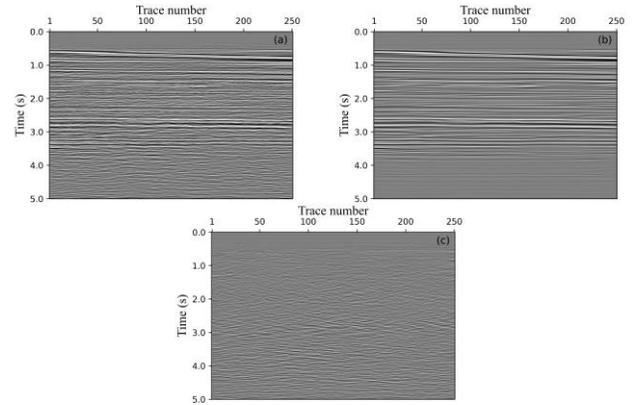

Fig. 7. Post-stack seismic data. (a) Noisy seismic data. (b) Denoised result by SSDCN. (c) Removed noise by SSDCN.